\documentclass[twocolumn,aps,prd,amssymb,nofootinbib,eqsecnum]{revtex4}
\pdfoutput=1
\usepackage{epsfig}
\usepackage{amsmath}
\usepackage{amssymb}
\usepackage{graphicx}
\newcommand{\vect}[1]{\boldsymbol{#1}}

\begin{document}
\title{Vacuum Instability in Chern-Simons Gravity}
\author{Sergei Dyda$^1$, \'Eanna \'E. Flanagan$^{1,2}$ and Marc Kamionkowski$^3$}

\affiliation{$^1$Center for Radiophysics and Space Research, Cornell
University, Ithaca, NY 14853}
\affiliation{$^2$Laboratory of Elementary Particle Physics, Cornell University, Ithaca, NY 14853}
\affiliation{$^3$Department of Physics and Astronomy, Johns
     Hopkins University, 3400 N.\ Charles St., Baltimore, MD 21218}

%\date{\today}
\def\no{\noindent}

\begin{abstract}

We explore perturbations about a Friedmann-Robertson-Walker background
%with a non-vanishing cosmological Chern-Simons scalar field
in Chern-Simons gravity. At large momenta one of the two circularly
polarized tensor modes becomes ghostlike.  We argue that nevertheless
the
theory does not exhibit classical runaway solutions, except possibly
in the relativistic nonlinear regime.  However, the ghost modes cause
the vacuum state to be quantum mechanically unstable, with a decay rate
that is naively infinite.  The decay rate can be made finite only if one
interprets the theory as an effective quantum field theory valid up to
some momentum cutoff $\Lambda$, which violates Lorentz invariance.
By demanding that the energy density in photons created by vacuum
decay over the lifetime of the Universe not violate observational
bounds, we derive strong constraints on the two dimensional parameter
space of the theory, consisting of the cutoff $\Lambda$ and the
Chern-Simons mass.
\end{abstract}

\maketitle

\def\be{\begin{equation}}
\def\ee{\end{equation}}
\def\bfx{{\bf x}}
\def\bfq{{\bf q}}
\def\bfJ{{\bf J}}
\def\bfk{{\bf k}}
\def\bfv{{\bf v}}
\def\bfzero{{\bf 0}}
\def\bfPsi{\mbox{\boldmath $\Psi$}}
\def\bfpsi{\mbox{\boldmath $\psi$}}
\def\bfnabla{\mbox{\boldmath $\nabla$}}
\def\bfsigma{\mbox{\boldmath $\sigma$}}
\def\bfomega{\mbox{\boldmath $\omega$}}
\def\bfOmega{\mbox{\boldmath $\Omega$}}
\def\bfTheta{\mbox{\boldmath $\Theta$}}
\def\bfcalJ{\mbox{\boldmath ${\cal J}$}}
\def\bea{\begin{eqnarray}}
\def\eea{\end{eqnarray}}
\def\nn{\nonumber}
\def\tt{{\tilde t}}
\newcommand{\bes}{\begin{subequations}}
\newcommand{\ees}{\end{subequations}}

\section{Introduction and Summary}

General relativity has held up well to
various tests over the years from experiments and astronomical
observations \cite{GRreview}, and is considered a pillar of standard
cosmology.
However, it is interesting to consider modifications to the theory,
particularly in light of the observed acceleration of the Universe
\cite{Skordis}.  One useful approach is
to think of Einstein gravity as an effective field theory,
and consider higher order corrections to the Einstein-Hilbert action,
either involving the metric alone or involving an additional posited
scalar field.
The goal then becomes to calculate the corrections to general
relativity arising from these higher order terms, and using
experiments to set bounds on the couplings parameterizing them.

One such extension to general relativity is
Chern-Simons gravity
\cite{Jackiw,CSreview}, where one assumes the existence of a
scalar field $\vartheta$ coupled to gravity through a parity violating
term. The
theory is described by
the action\footnote{Throughout this paper we will restrict attention to
the theory (\ref{first}) in which the scalar field is dynamical; we
will not consider the ``non-dynamical'' version of the theory in which
the kinetic term for the scalar field is absent
\protect{\cite{CSreview}}.  We note however that our derivation of the action (\protect{\ref{eq:quadratic}}) of the ghost graviton modes is valid for the non-dynamical theory, since the derivation does not involve any perturbations to the scalar field.}
\begin{equation}\label{first}
 S = S_{EH} + S_{CS} + S_{\vartheta} + S_{\rm mat},
\end{equation}
where the various terms are respectively the Einstein-Hilbert term
\begin{subequations}
\begin{equation}
S_{EH} = \frac{1}{2}m^2_{\rm p} \int d^4x \sqrt{-g} R,
\label{eq:EinsteinHilbert}
\end{equation}
the Chern-Simons term
\begin{equation}
S_{CS} = \frac{1}{4} \alpha \int d^4x \sqrt{-g} \ \vartheta \ ^*RR ,
\label{eq:CS}\\
\end{equation}
the scalar term
\begin{equation}
 S_{\vartheta} = - \frac{1}{2} \int d^4x \sqrt{-g} \ [\nabla_a \vartheta \nabla^a \vartheta + 2V(\vartheta)],
\label{eq:scalar}
\end{equation}
\end{subequations}
and $S_{\rm mat}$ describes any other matter present. In these
expressions $m^2_{\rm p} = (8\pi G)^{-1}$ is the square of the reduced
Planck mass,
$g$ is the determinant of the metric,
$R$ is the Ricci scalar and $\alpha$ a coupling constant with
dimensions of inverse mass.   Here and throughout we use units with
$\hbar = c = 1$.  Also
$V(\vartheta)$ is an
arbitrary potential
and the Pontryagin density is defined as
\begin{equation}
 ^*RR = \frac{1}{2} \epsilon^{cdef}R^a_{\ bef}R^b_{\ acd},
\end{equation}
where $\epsilon^{cdef}$ is the four dimensional Levi-Civita tensor. For purposes of the present discussion, we will assume that matter is minimally coupled to the metric.

Consider now the dynamics of Chern-Simons gravity in perturbation
theory about
a Friedmann-Robertson-Walker (FRW)
cosmological background.
Now in the limit
$\vartheta =$ constant, the Chern-Simons term (\ref{eq:CS}) reduces to
a surface term and we recover Einstein's equations for the space-time
dynamics.  Therefore, in the effective field theory that describes the
perturbations, the operators that arise from the Chern-Simons term
must be
suppressed by a mass scale that is related to the derivative of the
background scalar field.  This mass scale is called the Chern-Simons
mass $m_{\rm cs}$, and is defined in the FRW context by \cite{solar}
\begin{equation}
 m_{\rm cs} \equiv \frac{m^2_{\rm p}}{\alpha \dot{\vartheta}},
\label{eq:CSscale}
\end{equation}
where the dot denotes a derivative with respect to time.  General
relativity is recovered in the limit $m_{\rm cs} \rightarrow
\infty$ for linearized tensor perturbations.
Because it is the
Chern-Simons mass that enters into
equations describing observables for linear perturbations, we choose
to constrain it, rather
than the more fundamental coupling $\alpha$ that appears in the
action (\ref{eq:CS}).
Also, since the background cosmological
solution $\vartheta(t)$ need not be a linear function of time, the
Chern-Simons mass will in general be a function of time or of
redshift; we will focus in this paper on its value $m_{\rm cs} =
m_{\rm cs}(t_0)$ today.

Past work constraining Chern-Simons gravity has utilized Solar System
and binary pulsar tests of general relativity.
Measurement of Lense-Thirring precession by LAGEOS give the bound
\cite{solar}
\begin{equation}
m_{\rm cs} \gtrsim 2 \times 10^{-13} \, {\rm eV}.
\end{equation}
A bound $10^{11}$ times stronger has been claimed from binary pulsar studies \cite{pulsar},
but the validity of this result has been questioned
\cite{pulsarcriticism}, and a corrected bound\footnote{Although this
  bound was derived in the context of the non-dynamical Chern-Simons
  theory, without the kinetic term for the scalar field, it is also
  valid for the dynamical theory when the cosmological
  background solution $\vartheta(\eta)$ is nonzero, since the derivation
  is based on the
  modified gravitational Amp\`ere equation which is valid in the
  dynamical theory \cite{solar}.} from binary pulsars is
\cite{pulsarcriticism}
\begin{equation}
m_{\rm cs} \gtrsim 5 \times 10^{-10} \, {\rm eV}.
\label{fixed}
\end{equation}

In this paper we study the vacuum stability of
Friedmann-Robertson-Walker
(FRW) solutions in Chern-Simons gravity as a function of the
Chern-Simons mass parameter.
In Sec.\ \ref{sec:ghosts} we consider tensor perturbations to the FRW
metric. We show that for spatial momenta above the Chern-Simons mass
scale, one of the two
polarization modes is ghostlike
and can decay to radiation.  Requiring that the radiation produced
over the lifetime of the Universe not exceed observational bounds
allows us to constraint the parameters of the theory, which we do in
Sec.\ \ref{sec:vacuumdecay}.  Finally in Section \ref{sec:nonpert} we
argue that
the theory does not exhibit classical runaway solutions,
despite the existence of ghostlike modes,
except possibly in the relativistic nonlinear regime.

\section{The existence of ghost graviton modes}
\label{sec:ghosts}

In Chern-Simons gravity, we assume that the background cosmological model is
homogeneous and isotropic as in general relativity.  The metric is of the form
\be
ds^2 = a(\eta)^2 [- d\eta^2 + \delta_{ij} d\chi^i d\chi^j],
\ee
where $a(\eta)$ is the scale factor, $\eta$ is conformal time, and $\chi^i$ are comoving coordinates, and the scalar field is $\vartheta = \vartheta(\eta)$.
[For simplicity we assume spatial flatness.]
The corresponding equations of motion do not contain any contribution from
the Chern-Simons term, since the FRW symmetries
lead to a vanishing Pontryagin density \cite{FRW}.  We therefore
obtain the usual Friedmann equations with a scalar field
\be
3 m_{\rm p}^2 \frac{a^{\prime\,2}}{a^4} = \rho_m +
\frac{\vartheta^{\prime\,2}}{2 a^2} + V(\vartheta),
\label{fr1}
\ee
and
\be
\vartheta^{\prime\prime} + 2 \frac{a^\prime}{a} \vartheta^\prime + a^2
V'(\vartheta) = 0,
\label{fr2}
\ee
where primes denote derivatives with respect to $\eta$ and $\rho_m$ is
the matter density.

We now fix a solution $a(\eta)$, $\vartheta(\eta)$ of the background
equations (\ref{fr1}) and (\ref{fr2}).  We assume that $\vartheta'(\eta)$
is not identically vanishing.  While there may exist solutions with
$\vartheta'(\eta) = 0$, if the potential has local minima,
these will not be generic.
Consider now linear perturbations about such a solution.
As in general relativity, the symmetries of the background solution
guarantee that perturbations can be decomposed into scalar,
vector and tensor modes.
In this paper we will focus on tensor
perturbation modes, for which the perturbation to the scalar field
vanishes and the metric takes the form
\begin{equation}
 ds^2 = a^2(\eta) [-d\eta^2 + (\delta_{ij} + h_{ij})d\chi^id\chi^j].
\end{equation}
We adopt the transverse and traceless
gauge conditions, $h^i_{\ i} = 0, \ \partial_ih^{ij} = 0 $. Expanding the terms in the action (\ref{first}) to quadratic order and simplifying using the background equations (\ref{fr1}) and (\ref{fr2}) yields \cite{birefringent}
\begin{eqnarray}
S &=& \frac{1}{8} \int d\eta \ d^{\: 3}\chi \bigg[m_{\rm p}^2 a^2(\eta) (h^{i}_{\ j, \eta} h^{j}_{\ i, \eta } - h^{i}_{\ j, k } h^{j\hspace{.15cm} ,k}_{\ i}) \nonumber \\
 &&- \ \alpha \vartheta_{,\eta} \ \epsilon^{ijk}  \ (h^{q}_{\ i,\eta } h_{kq,j \eta } - h^{q\hspace{.15cm} ,r}_{\ i}h_{kq,rj}) \bigg] \nonumber \\
&& + \ O(h^3),
\label{eq:quadratic}
\end{eqnarray}
where $\epsilon^{ijk}$ is the Levi-Cevita symbol. We rewrite this action in terms of the Fourier transform of the metric perturbation, which is defined by
\begin{equation}
 h_{ij}(\eta,\vect{\chi}) = \int d^3k \ \tilde{h}_{ij}(\eta,\vect{k}) \ e^{i \vect{k} \cdot \vect{\chi}},
\end{equation}
where $\vect{k}$ is the comoving wavevector.

As noted by Refs.\ \cite{birefringent,Alexander:2004us}, the dynamics
is simplest when expressed in terms of a circular polarization basis.
We define the left and right circular polarization modes $\tilde{h}_{A \vect{k}}(\eta)$ by
\begin{equation}
 \tilde{h}_{ij}(\eta, \vect{k}) = \sum\limits_{A = L,R} \tilde{h}_{A \vect{k}}(\eta)\: e^{A}_{\ ij}(\vect{n}).
\end{equation}
where $\vect{n} = \vect{k}/k $ is a unit vector in the direction of
propagation, $k \equiv |\vect{k}|$, and
the polarization tensors $e^{A}_{\ ij}(\vect{n})$ satisfy the conditions
\begin{subequations}
\begin{equation}
 e^{A}_{\ ij}(e^{B}_{\ ij})^* = 2\delta^{AB},
\end{equation}
\begin{equation}
 n_i \epsilon^{ijk} e_{kl}^{A} = i \lambda_{A} (e^{j}_{\ l})^{A},
\end{equation}
\end{subequations}
with $\lambda_R = +1$ and $\lambda_L = -1$. The action can
now be written as
\begin{eqnarray}
S &=& \frac{m_{\rm p}^2}{4} \int d\eta \ d^3k  \sum\limits_{A = L,R} a^2(\eta) \nonumber \\
&& \left[1 +  \frac{\lambda_{A} k}{a^2(\eta)}  \frac{\alpha \vartheta_{,\eta}}{m_{\rm p}^2}\right]\left[|\tilde{h}_{A \vect{k},\eta}|^2 - k^2|\tilde{h}_{A \vect{k}}|^2 \right].
\label{eq:GWsolution}
\end{eqnarray}
This can be recast using the Chern-Simons mass scale
(\ref{eq:CSscale}) as
\begin{eqnarray}
  S &=& \frac{m_{\rm p}^2}{4} \int d\eta \ d^3k \sum\limits_{A = L,R} a^2(\eta) \nonumber \\
 && \left[ 1 + \lambda_{A} \frac{k_{\text{phy}}}{m_{\rm cs}}\right] \left[|\tilde{h}_{A \vect{k},\eta}|^2 - k^2|\tilde{h}_{A \vect{k}}|^2\right],
\label{eq:ghost}
\end{eqnarray}
where $k_{\text{phy}} = k/a$ is the physical wavenumber. The action
(\ref{eq:ghost}) is the usual action for tensor perturbations in FRW,
except for the momentum dependent correction factor
$(1 + \lambda_{A} k_{\text{phy}}/m_{\rm cs})$.
This factor becomes negative for the left handed polarization modes
when $k_{\text{phy}} \geq m_{\rm cs}$, giving rise to a kinetic term
with the wrong sign, i.e. a ghost mode.

The action (\ref{eq:ghost}) can be simplified by
changing the normalization of the graviton modes to attain canonical
normalization.  We define
\be
\varepsilon_A(k) = {\rm sgn} \left( 1 + \frac{\lambda_A k_{\rm phys} }{m_{\rm cs}} \right),
\ee
which is $+1$ for normal modes and $-1$ for ghost modes.
We define the mass scale
\be
m_* = m_{\rm p} \sqrt{ \left|  1 + \frac{\lambda_A k_{\rm phys} }{m_{\rm cs}} \right| },
\ee
and the canonically normalized graviton field modes
\be
{\tilde h}^{\rm can}_{A\vect{k}} =
m_* {\tilde h}_{A\vect{k}}.
\ee
The action (\ref{eq:ghost}) can now be written as
\begin{eqnarray}
  S &=& \frac{1}{4} \int d\eta \ d^3k \sum\limits_{A = L,R} a^2(\eta) \varepsilon_A(k) \nonumber \\
 && \left[\left|\tilde{h}^{\rm can}_{A \vect{k},\eta}
- (\ln m_*)_{,\eta} \tilde{h}^{\rm can}_{A \vect{k}}\right|^2 -
   k^2|\tilde{h}^{\rm can}_{A \vect{k}}|^2\right].
\label{eq:ghost1}
\end{eqnarray}

Note that from a classical point of view the existence of ghost modes
does not necessarily imply any inconsistency of the theory.
The Hamiltonian of the theory may or may not be unbounded below;
addressing this question would require a nonlinear analysis beyond
the scope of this paper.  Also it is not known at present whether the
theory possesses a well posed initial value formulation: the
sign flip at $k_{\rm phys} = m_{\rm cs}$
may be a hint that it does not.
However, the ghost modes are a significant problem
when quantum mechanical effects are taken into account, as we
discuss in the next section.

%One might also expect that the ghost graviton modes would be
%associated with fluxes of negative energy.  Rather surprisingly, this
%is not so; a computation of the effective stress energy tensor for
%gravitational waves at future null infinity shows that all the
%graviton modes carry positive energy \cite{Stein:2010pn}.

\section{Constraints from vacuum decay}
\label{sec:vacuumdecay}

We now specialize to perturbation modes today which are deep inside
the horizon, that is, $k \gg H_0$, where $H_0 = a_{,\eta} / a^2$ is the
Hubble parameter.  We also assume that $m_{\rm cs} \gg H_0$.  In this
limit, we can neglect in the action (\ref{eq:ghost1})
the time dependence of the prefactor $a(\eta)^2$,
and also the term proportional to the time derivative of
$m_*$.
The result is just the standard
action for graviton modes in Minkowski spacetime, except for the sign
flip for the ghost modes\footnote{We have omitted the action
describing the interaction of the graviton field modes with matter
fields.  This interaction takes the standard form when written in
terms of ${\tilde h}_{A\vect{k}}$, but acquires correction factors of
 $m_{\rm p}/m_*$ when written in terms of
${\tilde h}^{\rm can}_{A\vect{k}}$.}.

Because of this sign flip, decay of the vacuum of
the theory is kinematically allowed.
The vacuum decay rate per unit volume $\Gamma$ is naively infinite,
because of the
infinite phase space available for the decay products that arises from
Lorentz invariance\footnote{The full theory of the
  perturbations about FRW, including coupling to matter, does violate
  Lorentz invariance because of the $k_{\rm phys}$ dependence of the factor $1 +
  \lambda_A k_{\rm phys}/m_{\rm cs}$.  However this violation does not
  affect the accessible volume of phase space.}.
Rather than ruling out
the theory outright because of the presence of ghost modes, we adopt
the viewpoint that the action (\ref{eq:ghost}) defines an effective
quantum field theory as in Refs.\
\cite{CSreview,Yagi:2011xp,Yagi:2012ya}, with some effective cutoff
$\Lambda$ on the physical
wavenumber $k_{\rm phys}$ in cosmological rest frame. Thus, our
cutoff explicitly
violates
Lorentz invariance; such a violation is inevitable if one wants to
obtain a finite
vacuum decay rate.  Our viewpoint and treatment follow similar
analyses of scalar ghost fields in cosmological models with equation
of state parameter $w$ that satisfies $w < -1$ \cite{Carroll,Cline}.
As in those
analyses, we
will find that stringent constraints on the parameters of the theory
%(here the cutoff $\Lambda$ and the Chern-Simons mass $m_{\rm cs}$)
can be obtained by demanding that the total number of photons produced
from vacuum decay over the lifetime of the Universe not be in conflict
with observations.

Our theory is now parameterized by two parameters, the cutoff
$\Lambda$ and the Chern-Simons mass $m_{\rm cs}$.  Now since the ghost
modes arise only for momenta $k_{\rm phys}$ satisfying $k_{\rm phys} >
m_{\rm cs}$, there will be no ghost modes in the region $m_{\rm cs} >
\Lambda$ of parameter space.
Therefore our arguments about vacuum decay do not constrain
that region of parameter space; we focus from now on on the
complementary region $m_{\rm cs} < \Lambda$ (see Fig.\ \ref{fig:phase
  space}). We note that many papers on Chern-Simons gravity
implicitly work in this regime $m_{\rm cs} < \Lambda$, since they
involve effects arising on scales $k_{\rm phys} \sim m_{\rm cs}$,
assumed to be within the domain of validity of the theory.
We also note that the regime $m_{\rm cs} \alt \Lambda$ is disfavored
by naturalness arguments, i.e., it requires considerable fine tuning of
the Lagrangian (see Appendix \ref{sec:natural}).

If the vacuum perturbatively decays to stable particles, we can estimate the number of particles produced over the lifetime of the Universe from such a process. Given observational bounds on the energy density
in this particle species, we can then constrain the decay rate. The
strongest bounds will therefore come from production of particles whose
energy density has been reliably measured, and which
can be produced at a low order in perturbation theory. We argue that
the photon is the best such candidate.

\subsection{Derivation of photon energy spectrum}

\begin{figure}[t!]
\vspace{-1cm}
\centering
    \includegraphics[width=0.4\textwidth]{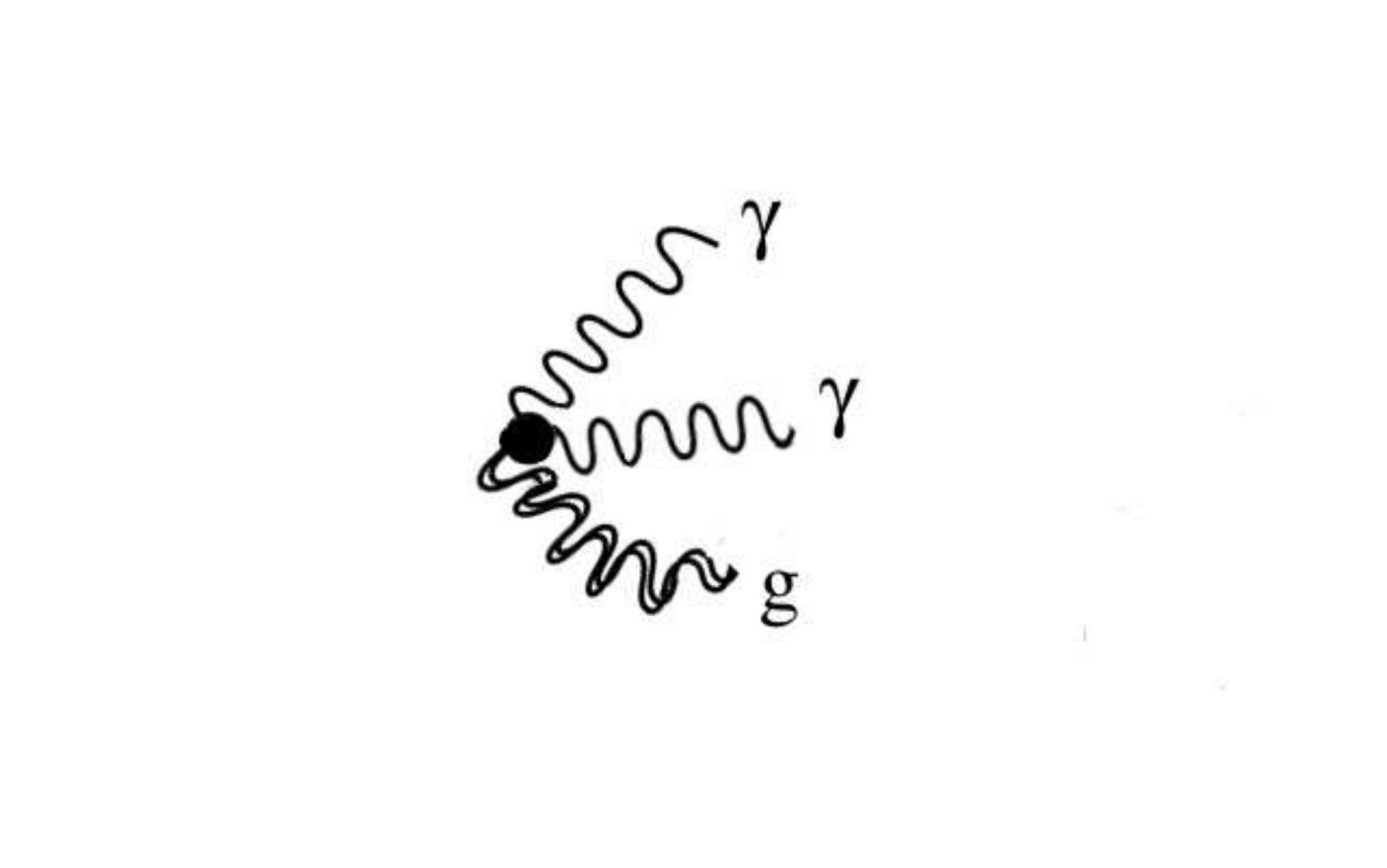}
  \vspace{-1cm}
\caption{Spontaneous production of photons and ghost gravitons from vacuum}
\label{fig:graviton}
\end{figure}

We next make an order of magnitude estimate of the decay rate $\Gamma$
per unit volume to photons in the regime $m_{\rm cs} < \Lambda$.
The action for electromagnetism contains
an interaction of
the form
\begin{equation}
 S_{\rm int} \sim  \int d^4x \ h \, (\partial A)^2 \sim
\frac{1}{m_*} \int d^4x \ h^{\rm can} \, (\partial A)^2,
\label{eq:interaction}
\end{equation}
where $A^a$ is the 4-vector potential, $h$ is the metric
perturbation, and $h^{\rm can}$ is the canonically normalized version.
Because the graviton is ghostlike,
the process
\begin{equation}
 0 \rightarrow g \gamma \gamma
\end{equation}
is kinematically
allowed, where $g$ is a left polarized graviton and $\gamma$ is a photon
(Fig. \ref{fig:graviton}). Hence graviton ghosts can decay to photons
at first order in perturbation theory.

Next, the coefficient $1/m_*$ in the interaction
(\ref{eq:interaction}) depends on the wavenumber $k$.
However,
for the purposes of our order of magnitude estimate, it will be
sufficient to evaluate this coefficient at $k_{\rm phys} \sim
\Lambda$, since most
of the decays will be at $k_{\rm phys} \sim \Lambda$.  Therefore we
can treat $m_*$ as a constant,
\be
m_* \sim m_{\rm p} \sqrt{ \frac{\Lambda}{m_{\rm cs}}},
\ee
where we have used $m_{\rm cs} < \Lambda$ and specialized to ghost
modes.
The decay rate $\Gamma$ per unit time per unit volume
must be proportional to the square of the coefficient
of the operator, so $\Gamma \propto 1/m_*^2$.
The constant of proportionality in this relation must be
some function of $\Lambda$, and from dimensional
analysis it now follows that\footnote{This computation breaks down in the limit $m_{\rm cs} \to \Lambda$, in which the volume of the region in phase space containing unstable modes shrinks to zero.  The computation is valid to within a factor of order unity whenever $\Lambda/m_{\rm cs} -1 \agt O(1)$.}

\begin{equation}
 \Gamma_{0\rightarrow g\gamma \gamma} \sim \frac{\Lambda^6}{m_*^2}
 \sim \frac{m_{\rm cs}\Lambda^5}{m_{\rm p}^2},
\end{equation}
since $\Gamma$ has dimension (mass)$^4$ in units with $\hbar = c =1$.

Consider now the production of photons by this mechanism over the
lifetime of the Universe.  We make the idealization that at any epoch,
all the photons produced have an energy of exactly $\Lambda$.  Then
the evolution of the number $n(k,t)$ of photons per unit logarithmic
wavenumber $k$ per proper volume is given by
\be
\frac{1}{a^3} \frac{d}{dt} \left( a^3 n\right) = \Gamma \Lambda
\delta(k/a - \Lambda),
\ee
where $a$ is the scale factor with $a=1$ today.  Solving this equation
gives for the spectrum today
\be
n(k) = \Theta(\Lambda - k) \frac{\Gamma a_*^3}{H_*},
\label{gen}
\ee
where $\Theta$ is the step function, $H = {\dot a}/a$ is the Hubble
parameter, and $a_*(k)$ and $H_*(k)$ are the values of $a$ and $H$
at $k/a = \Lambda$.
The result (\ref{gen}) is easy to understand:
the factor of $1/H_*$ is
the length of time during which photons of present-day energy $\sim k$ are produced, and the factor of $a_*^3$ is
the volume expansion factor
since then.  For a $\Lambda$CDM cosmology with $H^2 = H_0^2 ( \Omega_M
/a^3 + 1 - \Omega_M)$ the spectrum becomes
\be
n(k) = \Theta(\Lambda - k ) \frac{\Gamma}{H_0} \left[ \Omega_M \left(
    \frac{\Lambda}{k}\right)^9 + (1-\Omega_M) \left( \frac{\Lambda}{k}
  \right)^6 \right]^{-1/2}.
\ee
It follows that the energy density per unit logarithmic
wavenumber $dE/d^3x d\ln k \sim k n(k)$ is peaked at $k \sim \Lambda$
with a peak value of
\be
\frac{d E}{d^3x d\ln k} \sim \frac{\Gamma \Lambda}{H_0} \sim
\frac{m_{\rm cs} \Lambda^6}{m_{\rm p}^2 H_0}.
\label{eee}
\ee

Finally, we note that the above derivation implicitly assumed that the
Chern-Simons mass $m_{\rm cs}$ is constant.  In reality $m_{\rm cs}$
is a function of redshift or of time, since the background
cosmological solution $\vartheta(t)$ is time dependent.
However, our derivation shows most of the photons are produced at low
redshift, $z \alt 1$.  Since the background cosmological solution
evolves on the Hubble timescale, the fractional change in the
Chern-Simons mass out to redshifts of order unity is of order
unity.  Therefore the result (\ref{eee}) is still valid when
taking the evolution of $m_{\rm cs}$ into account, up to a correction
factor of order unity.

\subsection{Comparison of photon energy spectrum with observations}

We now compare the prediction (\ref{eee}) with observational data,
in two different ways.  First, as long as $\Lambda \agt H_0$, the
energy density (\ref{eee}) will contribute to the expansion of the
Universe.  Since observations of the expansion history tell us that
the Universe is not radiation dominated today, we obtain as a very
conservative upper bound that
\be
\frac{dE}{d^3x d\ln k} \alt m_{\rm p}^2 H_0^2.
\ee
It follows that
\be
m_{\rm cs} \Lambda^6 \alt m_{\rm p}^4 H_0^3 \sim (38 \, {\rm eV})^7
\ \ \ {\rm for } \ \ 10^{-33} \, {\rm eV} \alt \Lambda,
\label{eq:photonconstraint1}
\ee
where we have used  $m_{\rm p} = 2.4
\times 10^{27} \, {\rm eV}$ and $H_0 = 1.5 \times 10^{-33} \, {\rm eV}$.

\begin{figure}[t!]
%\vspace{-3.7cm}
\centering
    \includegraphics[width=8.0cm]{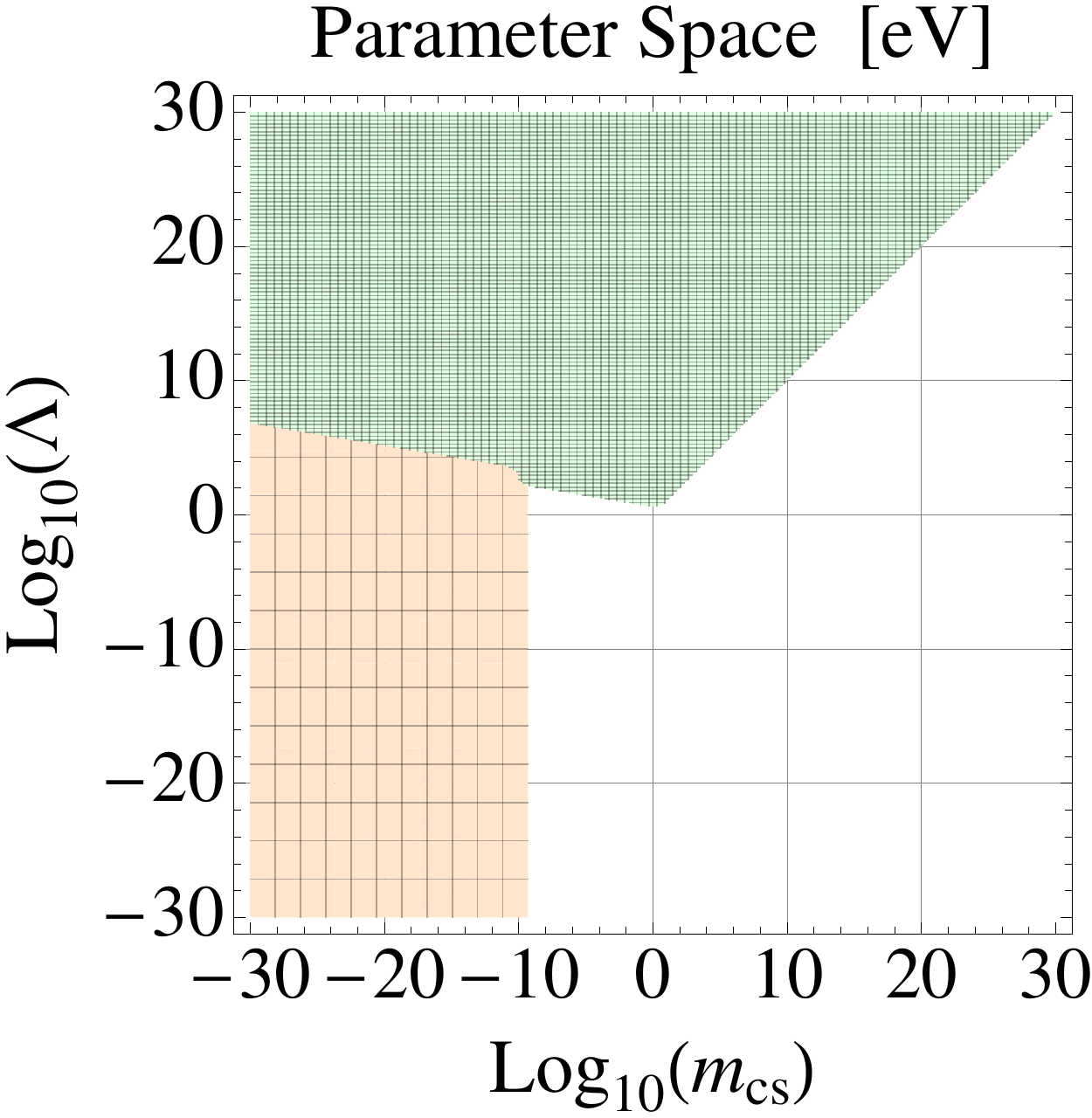}
% \vspace{-1.7cm}
\caption{Effective field theory parameter space consisting of the
cutoff scale $\Lambda$ and the Chern-Simons mass $m_{\rm
    cs}$, in eV.  The lightly shaded region is excluded from binary pulsar
observations \protect{\cite{pulsarcriticism}}, and the
darkly shaded region is excluded by consideration of vacuum decay.}
\label{fig:phase space}
\end{figure}

Second, a more stringent constraint can be obtained by using direct
observations of photons, albeit within a finite range of cutoffs
$\Lambda$.
Observations in
various wavelength bands from radio to gamma rays gives the upper
bound\footnote{This upper bound is exceeded by about two
orders of magnitude by the CMB, but the CMB is known to be thermal to
about 1 part in $10^5$, so the bound effectively holds.}
 on the background radiation spectrum
\be
\frac{dE}{d^3x d \ln k} \alt (2.5 \times 10^{-5} \, {\rm eV})^4
\ee
for $10^{-10} \, {\rm eV} \alt k \alt 10 \, {\rm GeV}$ \cite{Henry,low}.
Combining this with the prediction (\ref{eee}) yields the constraint
\be
m_{\rm cs} \Lambda^6 \alt (3.3 \, {\rm eV})^7
\ \ \ {\rm for } \ \ 10^{-10} \, {\rm eV} \alt \Lambda \alt 10 \, {\rm GeV}.
\label{eq:photonconstraint}
\ee
Figure \ref{fig:phase space} shows the constraints (\ref{fixed}) and
(\ref{eq:photonconstraint1}), (\ref{eq:photonconstraint}) on the
$(m_{\rm cs},\Lambda)$ parameter
space.
It can be seen that the combined constraints rule out a large part of
the region in parameter space with
$\Lambda > m_{\rm cs}$ in which interesting
Chern-Simons phenomenology occurs, except for the window
$10^{-10} \, {\rm eV} \alt m_{\rm cs} \alt 3\, {\rm eV}$.

\label{sec:pert}

We note that our constraint  (\ref{eq:photonconstraint1})
 on the parameter space extends up
to arbitrarily large values of $m_{\rm cs}$, even though the theory
ostensibly reduces to general relativity in this limit to linear order
in perturbation theory.
This is because for any value of $m_{\rm cs}$, there exist modes with
$k_{\rm phys} \sim m_{\rm cs}$ for which the Chern-Simons correction
factor in Eq.\ (\ref{eq:ghost}) is of order unity.
Even if $m_{cs}$ is arbitrarily large, the high energy modes
above this scale are unstable and the vacuum decays.  The high energy
photons produced can then give an observable contribution to the
expansion rate of the Universe.

\subsection{Vacuum decay to gravitons}

One can also compute the constraints obtained from vacuum decay to
gravitons, in which one ghost graviton and two normal gravitons are
produced.   The interaction Lagrangian is
\be
{\cal L}_{\rm int} \sim m_{\rm p}^2 h (\partial h)^2 \sim {m_{\rm p}^2
  \over m_*^3} h_{\rm can} (\partial h_{\rm can})^2.
\ee
The analysis now proceeds as before, with the resulting energy density
in gravitons given by
\be
\Omega_{\rm gw} \sim {1 \over m_{\rm p}^2 H_0^2} {d E \over d^3 x d\ln
  k} \sim {\Lambda^4 m_{\rm cs}^3 \over m_{\rm p}^4 H_0^3}
\label{eq:gravitons}
\ee
at $k \sim \Lambda$.  Now observations of the cosmic microwave
background give an integrated upper bound
$\int d\ln k \Omega_{\rm gw} \alt 10^{-5}$ for $k \agt 10^{-30}\,{\rm
  eV}$ \cite{cmb}.  However this bound applies only to the gravitons
produced before recombination, whose energy density will be smaller than
the estimate (\ref{eq:gravitons}) by a factor of the ratio of the age of the Universe at recombination to the age of the Universe today.
This factor is approximately $z_{\rm
    rec}^{-3/2}$, where $z_{\rm rec} \sim 1000$ is the redshift of
recombination.  Also for gravitons produced before recombination,
the Chern-Simons mass $m_{\rm cs} = m_{\rm cs}(t_0)$ in Eq.\
(\ref{eq:gravitons}) should be replaced by ${\tilde m}_{\rm cs} =
m_{\rm cs}(t_{\rm rec})$, the value at recombination.
We therefore obtain
the constraint
\be
{\tilde m}_{\rm cs}^3 \Lambda^4 \alt z_{\rm rec}^{3/2} m_{\rm p}^4
H_0^3 \,
(10^{-5}) \sim (32 \, {\rm eV})^7
\label{eq:gravitonconstraint}
\ee
for $\Lambda \agt 10^{-30} \, {\rm eV}$.
The result (\ref{eq:gravitonconstraint}) is slightly weaker than the
constraint (\ref{eq:photonconstraint}) from photons, and applies
to a different version of the Chern-Simons mass.

Finally, one might also be able to place additional constraints on the
theory by considering the decay of the vacuum to two gravitons and one
quanta  of the scalar field, rather than to three gravitons.

\section{Classical Runaway Solutions}
\label{sec:nonpert}

We next turn to a different question, the possible existence of
classical runaway solutions that can occur in a theory whose
Hamiltonian is unbounded below.  We will argue that such solutions do
not arise in Chern-Simons gravity except possibly in the nonlinear
relativistic regime $h \sim 1$ where our analysis  is invalid anyway.

Generically, a Hamiltonian which is unbounded from below
will only exhibit runaway solutions in the regime where the
interaction terms in the Lagrangian are comparable to the negative
kinetic energy terms \cite{Carroll}.
We use this criterion to estimate the required occupation number of
a ghost graviton mode for such a runaway behavior to develop in our
system.

No runaway solutions will plague the effective field theory (\ref{eq:ghost}) because the ghost and non-ghost fields are decoupled to linear order.
However, interaction terms will appear if we expand the action (\ref{first}) to higher than quadratic order. Expanding up to quartic order we expect to find
interaction terms of the form
\begin{subequations}
\begin{equation}
S_3 \sim \int d^4x \: m_{\rm p}^2 \; k^2 h_L^2h_R + ...,
\end{equation}
\begin{equation}
S_4 \sim \int d^4x  \: m_{\rm p}^2 \; k^2 h_L^2h_R^2 + ...
\ .
\end{equation}
\end{subequations}
%Classical instabilities form when the contributions to the energy from interaction terms are of the same order of magnitude as the kinetic term \cite{Carroll}.
Consider a single localized wave packet mode with characteristic size
$\lambda$.  We would like to estimate the number of quanta
$N$ for an instability to form. The energy from the kinetic term is
\begin{equation}
E_K = \frac{1}{2} \int d^3x \ m_{\rm p}^2 \: \dot{h}^2 \sim \lambda m_{\rm p}^2 h^2 \sim \frac{N}{\lambda}.
\end{equation}
Therefore the field fluctuation is approximately
\begin{equation}
h \sim \frac{N^{1/2}}{\lambda m_{\rm p}}.
\label{eq:h}
\end{equation}
Each of the interaction terms will therefore contribute terms to the energy on the order of
\begin{subequations}
\begin{equation}
E_3 \sim \int d^3x \; m_{\rm p}^2 \: k^2 h^3 \sim  \frac{N^{3/2}}{\lambda^2 m_{\rm p}},
\end{equation}
\begin{equation}
E_4 \sim \int d^3x  \; m_{\rm p}^2 \: k^2 h^4 \sim   \frac{N^2}{\lambda^3 m_{\rm p}^2}.
\end{equation}
\end{subequations}
Requiring that the kinetic energy be less than the interaction energy
\begin{equation}
%\begin{align}
E_3 \gtrsim E_K,   \hspace{1.5cm}   E_4 \gtrsim E_K,
%\end{align}
\end{equation}
and taking the wavelength $\lambda \sim m_{\rm cs}^{-1}$ yields
\begin{equation}
N \geq \left(\frac{m_{\rm p}}{m_{\rm cs}}\right)^2 \gg 1.
\end{equation}
We deduce from (\ref{eq:h}) that in this regime
\be
h \gtrsim 1.
\ee
We conclude the ghost modes generated by perturbations to FRW
do not exhibit runaway solutions within the domain of validity of our analysis.

\section{Outlook}

We have shown that if the effective field theory cutoff $\Lambda$ is
above the Chern-Simons mass scale $m_{\rm cs}$, and if the
Chern-Simons mass is not infinite (the generic case), then
the vacuum of
Chern-Simons gravity is unstable and can decay to photons.
Our estimate of the photon production rate was used to set constraints
on the parameter space.  One might imagine improving this bound by
being less cavalier in the estimation of the decay rate and number
density. However, because the production rate goes like a high power
of the cutoff scale, orders of magnitude difference in the number
density yield very modest improvements in final constraints on
$\Lambda$ and $m_{\rm cs}$.
%One might also imagine using a different measurement to set the bounds,
%for instance cosmic microwave background polarization constraints, since the photons produced will all be left polarized. Again this would only yield modest
%improvements because of the high power of the cutoff.

Whether the full nonlinear theory has a Hamiltonian which is unbounded
below is an interesting open question.

We also note that it is possible that the decay of the vacuum
could be
a lot faster than indicated by our perturbative calculations, if it is
mediated by a non-perturbative process.  One might imagine a tunneling
process similar to the decay of false vacua in scalar field theories
via bubble nucleation.  In particular, a homogeneous unstable region
with small Chern-Simons might be expected to produce a stable endstate
consisting of a homogeneous region with a Chern-Simons mass larger
than the cutoff.  However any such transition could not proceed via
a spherically symmetric process, since the Chern-Simons term does
not contribute to the dynamics in spherical symmetry.

Finally, we note that our analysis does not constrain
the (non-generic) scenario where the background cosmological
Chern-Simons field is sitting at a minimum of the potential, which
corresponds to an infinite Chern-Simons mass.  In this scenario, no
ghostlike tensor modes appear.  Deviations from general relativity for
localized sources do arise, at nonlinear order in
perturbation theory, via the Pontryagin density source term in the
scalar field equation.  It may be possible to use tests of general
relativity to constrain the parameter $\alpha$ in the action
(\ref{eq:CS}) in
this scenario.

\section*{Acknowledgments}

EF and SD were supported by NSF
grants PHY-0757735 and PHY-1068541 and by NASA grant NNX11AI95G.
MK was supported by DoE SC-0008108 and NASA NNX12AE86G.
We thank Leo Stein, Nico Yunes and Kent Yagi for helpful conversations.
EF thanks the Theoretical Astrophysics Including Relativity Group at
Caltech, and the Department of Applied Mathematics and Theoretical
Physics at the University of Cambridge, for their hospitality as this
paper was being initiated.

\appendix
\section{Naturalness Argument}
\label{sec:natural}

In this appendix we shall argue that the regime $\Lambda \agt m_{\rm
  cs} $ of parameter space is disfavored by naturalness arguments,
that is, it requires considerable fine tuning of the Lagrangian.
We specialize in this appendix to the version of the theory without a
potential, $V(\vartheta)=0$.

We start by assuming that Chern-Simons gravity arises from an
effective field
theory.  In this framework
we seek a description of the physics only at energies $E$ small
compared to a cutoff scale $\Lambda$, and we assume that at these
energies the only relevant degrees of freedom are a metric $g_{ab}$
and a scalar field $\vartheta$, with the only symmetries being
diffeomorphism invariance and the shift symmetry
$\vartheta \to \vartheta + $ constant.  As is usual,
the theory at energies
$E \agt \Lambda$ is unknown, but is assumed to be sufficiently random
or generic (i.e.\ no additional symmetries are present) that the
coefficients of all nonrenormalizable operators in the low energy
theory can be estimated using dimensional analysis to within
coefficients of order unity.

This framework leads to the following action, consisting of the
Chern-Simons action discussed in the introduction together with a
series of correction terms that are suppressed by powers of $\Lambda$:
\begin{eqnarray}
S &=& \int d^4x \sqrt{-g} \left[ \frac{1}{2}
  m_{\rm p}^2 R - \frac{1}{2} (\nabla \vartheta)^2
+\frac{1}{4} \alpha \vartheta \ ^*RR \right] \nonumber \\
\mbox{} && + \int d^4 x \sqrt{-g} \left[ \frac{c_1}{\Lambda^4} (\nabla
  \vartheta)^4 + \ldots \right].
\label{eq:EFT}
\end{eqnarray}
Here $c_1$ is dimensionless and of order unity.
The coefficient $\alpha$ can be estimated by using
\be
R \sim (\nabla h)^2 + \nabla \nabla h \sim \frac{1}{m_{\rm p}^2} (\nabla h^{\rm
  can})^2 + \frac{1}{m_{\rm p}} \nabla \nabla h^{\rm can},
\ee
where $h$ is the metric perturbation and $h^{\rm can} = m_{\rm p} h$
is the canonically normalized version.  This yields
\be
\alpha \vartheta \ ^*RR
\sim \frac{\alpha}{m_{\rm p}^2} \vartheta (\nabla \nabla h^{\rm
  can})^2
\sim \frac{c_2}{\Lambda^3} \vartheta (\nabla \nabla h^{\rm can})^2
\ee
where the last equality follows from dimensional
analysis and $c_2$ is dimensionless and of order unity.  It
follows that
\be
\alpha = c_2 \frac{m_p^2}{\Lambda^3}.
\label{eq:alpha}
\ee
This relationship between the coefficient $\alpha$ and the cutoff
$\Lambda$ was previously derived using a different method in Ref.\
\cite{Yagi:2012ya}.

By combining the relation (\ref{eq:alpha}) with the definition
(\ref{eq:CSscale}) of the Chern-Simons mass we obtain
\be
m_{\rm cs} = \frac{\Lambda^3}{c_2 {\dot \vartheta}}.
\label{eq:cs2}
\ee
We now make use of the fact that the background cosmological solution
must lie within the domain of validity of the effective field theory,
which requires that
\be
\frac{c_1}{\Lambda^4} (\nabla \vartheta)^4 \ll (\nabla \vartheta)^2
\ee
or ${\dot \vartheta} \ll \Lambda^2 /\sqrt{c_1}$.  Combining this with
Eq.\ (\ref{eq:cs2}) gives
\be
m_{\rm cs} \gg \frac{\sqrt{c_1}}{c_2} \Lambda,
\ee
and so $m_{\rm cs} \alt \Lambda$ is disallowed unless $\sqrt{c_1}/c_2
\ll 1$, which would be a fine tuning.

%
% other possible style files
%\bibliographystyle{apsrev}
%\bibliographystyle{prl}
%

%\bibliographystyle{physrev}
%\bibliography{CS}

\begin{thebibliography}{10}

\bibitem{GRreview}
C.~M. {Will}, ``{The Confrontation between General Relativity and
  Experiment}'',
\newblock Living Reviews in Relativity {\bf 9}, 3 (2006).

\bibitem{Skordis}
T.~Clifton, P.~G. Ferreira, A.~Padilla, and C.~Skordis, ``{Modified Gravity and
  Cosmology}'',
\newblock Phys.Rept. {\bf 513}, 1 (2012).
%%CITATION = ARXIV:1106.2476;%%

\bibitem{CSreview}
S.~Alexander and N.~Yunes, ``{Chern-Simons Modified General Relativity}'',
\newblock Phys.Rept. {\bf 480}, 1 (2009).
%%CITATION = ARXIV:0907.2562;%%

\bibitem{Jackiw}
R.~{Jackiw} and S.-Y. {Pi}, ``{Chern-Simons modification of general
  relativity}'',
\newblock \prd {\bf 68}, 104012 (2003).

\bibitem{solar}
T.~L. Smith, A.~L. Erickcek, R.~R. Caldwell, and M.~Kamionkowski, ``{The
  Effects of Chern-Simons gravity on bodies orbiting the Earth}'',
\newblock Phys.Rev. {\bf D77}, 024015 (2008).
%%CITATION = ARXIV:0708.0001;%%

\bibitem{pulsar}
N.~Yunes and D.~N. Spergel, ``{Double-binary-pulsar test of Chern-Simons
  modified gravity}'',
\newblock \prd {\bf 80}, 042004 (2009).

\bibitem{pulsarcriticism}
Y.~Ali-Haimoud, ``{Revisiting the double-binary-pulsar probe of non-dynamical
  Chern-Simons gravity}'',
\newblock Phys.Rev. {\bf D83}, 124050 (2011).
%%CITATION = ARXIV:1105.0009;%%

\bibitem{FRW}
D.~Guarrera and A.~J. Hariton, ``{Papapetrou energy-momentum tensor for
  Chern-Simons modified gravity}'',
\newblock \prd {\bf 76}, 044011 (2007).

\bibitem{birefringent}
S.~Alexander and J.~Martin, ``{Birefringent gravitational waves and the
  consistency check of inflation}'',
\newblock \prd {\bf 71}, 063526 (2005).

\bibitem{Alexander:2004us}
S.~H.-S. Alexander, M.~E. Peskin, and M.~M. Sheikh-Jabbari, ``{Leptogenesis
  from gravity waves in models of inflation}'',
\newblock Phys.Rev.Lett. {\bf 96}, 081301 (2006).
%%CITATION = HEP-TH/0403069;%%

%\bibitem{Stein:2010pn}
%L.~C. Stein and N.~Yunes, ``{Effective Gravitational Wave Stress-energy Tensor
%  in Alternative Theories of Gravity}'',
%\newblock Phys.Rev. {\bf D83}, 064038 (2011).
%%CITATION = ARXIV:1012.3144;%%

\bibitem{Yagi:2011xp}
K.~Yagi, L.~C. Stein, N.~Yunes, and T.~Tanaka, ``{Post-Newtonian,
  Quasi-Circular Binary Inspirals in Quadratic Modified Gravity}'',
\newblock Phys.Rev. {\bf D85}, 064022 (2012).
%%CITATION = ARXIV:1110.5950;%%

\bibitem{Yagi:2012ya}
K.~Yagi, N.~Yunes, and T.~Tanaka, ``{Slowly Rotating Black Holes in Dynamical
  Chern-Simons Gravity: Deformation Quadratic in the Spin}'',
\newblock Phys.Rev. {\bf D86}, 044037 (2012).
%%CITATION = ARXIV:1206.6130;%%

\bibitem{Carroll}
S.~M. Carroll, M.~Hoffman, and M.~Trodden, ``{Can the dark energy equation - of
  - state parameter w be less than -1?}'',
\newblock Phys.Rev. {\bf D68}, 023509 (2003).
%%CITATION = ASTRO-PH/0301273;%%

\bibitem{Cline}
J.~M. Cline, S.~Jeon, and G.~D. Moore, ``{The Phantom menaced: Constraints on
  low-energy effective ghosts}'',
\newblock Phys.Rev. {\bf D70}, 043543 (2004).
%%CITATION = HEP-PH/0311312;%%

\bibitem{Henry}
R.~C. {Henry}, ``{Diffuse Background Radiation}'',
\newblock Astrophysical Journal Letters {\bf 516}, L49 (1999).

\bibitem{low}
J.-L. {Bougeret}, ``{Very low frequency radio astronomy}'',
\newblock Advances in Space Research {\bf 18}, 35 (1996).

\bibitem{cmb}
T.~L. Smith, E.~Pierpaoli, and M.~Kamionkowski, ``{A new cosmic microwave
  background constraint to primordial gravitational waves}'',
\newblock Phys.Rev.Lett. {\bf 97}, 021301 (2006).
%%CITATION = ASTRO-PH/0603144;%%

\end{thebibliography}

\end{document}